\documentclass[letterpaper, twocolumn,superscriptaddress, showpacs,preprintnumbers,amsmath,amssymb] {revtex4}
\usepackage{graphicx}
\usepackage{dcolumn}
\usepackage{bm}
\usepackage{hyperref}
\hypersetup{
    colorlinks,
    citecolor=red,
    filecolor=blue,
    linkcolor=blue,
    urlcolor=blue
}

\providecommand{\abs}[1]{\lvert#1\rvert}
\newcommand{\ket}[1]{\left| #1 \right>} 
\newcommand{\matrixel}[3]{\left< #1 \vphantom{#2#3} \right|
 #2 \left| #3 \vphantom{#1#2} \right>} 

\begin{document}

\title{Entanglement spreading in a many-body localized system}

\author{Arun Nanduri}
\affiliation{Chemistry Department, Princeton University, Princeton, NJ 08544, USA}

\author{Hyungwon Kim}

\affiliation{Physics Department, Princeton University, Princeton, NJ 08544, USA}

\author{David A. Huse}
\affiliation{Physics Department, Princeton University, Princeton, NJ 08544, USA}

\begin{abstract}
  Motivated by the findings of logarithmic spreading of entanglement in a many-body localized system,
  we more closely examine the spreading of entanglement in the fully many-body localized phase,
  where all many-body eigenstates are localized.
  Performing full diagonalizations of an XXZ spin model with random longitudinal fields,
  we identify two factors contributing to the spreading rate:
  the localization length ($\xi$), which depends on the disorder strength,
  and the final value of entanglement per spin ($s_\infty$), which primarily depends on the initial state.
  We find that the entanglement entropy grows with time as $\sim \xi \times s_\infty \log t$,
  providing support for the phenomenology of many-body localized systems recently proposed
  by Huse and Oganesyan [arXiv:1305.4915v1].
\end{abstract}

\pacs{75.10.Pq, 03.65.Ud}

\maketitle

\section{Introduction}
More than half a century ago, Anderson pointed out the possibility that diffusion or conduction could be absent
in an isolated quantum many-body system due to static disorder~\cite{anderson}.
Although this original paper did discuss an interacting system in highly-excited states,
most of the subsequent work on Anderson localization focused instead on either noninteracting particles (or waves)
or on the low temperature limit.
It was Basko {\it et al.}~\cite{basko} that re-ignited interest in
highly-excited interacting systems with strong disorder,
where they found strictly zero conductivity at non-zero temperature.
This phenomenon is now called `many-body localization'.  Even highly excited states can be many-body localized.
Recently, many-body localization has attracted much attention,
for it allows new types of quantum phase transitions in highly-excited states~\cite{vadim,pal,order,pekker}
and is able to protect certain forms of topological order~\cite{order,bauer,chandran}.

While there is no DC transport of local observables in a many-body localized system \cite{basko},
it has been observed that entanglement
can still spread over the entire system, with the size of entangled regions growing with time $t$ as
$\sim\log (t)$~\cite{znidaric,bardarson}.
Several authors have examined this logarithmic dependence,
showing that it is a consequence of dephasing
due to the interactions~\cite{spa1,vosk,HO}.  Indeed, Refs.~\cite{znidaric,bardarson,spa1} report no growth of entanglement in the non-interacting case, and Ref.~\cite{pino} reports that entanglement entropy can grow faster in the presence of long-range interactions.
Here we restrict our consideration to only short-range interactions.
In particular, Ref.~\cite{HO} introduces a phenomenology of many-body localization of (short-range) interacting spins,
which naturally explains that the bipartite entanglement $S(t)$ of a one-dimensional many-body localized system initialized in a pure product state grows as
\begin{align}\label{entanglement}
S(t) \sim \xi s_\infty \log (t) ~,
\end{align}
where $\xi$ is a localization length (defined below) which primarily depends on the interaction and disorder strengths,
and $s_{\infty}$ is the final value of entanglement per spin~\cite{saturation},
which can strongly depend on the initial state.

In this paper, we examine the spreading of entanglement in detail
by using exact diagonalization of the random-field XXZ Hamiltonian. 
Varying the disorder strength and initial conditions,
we study how the spreading rate changes and thereby find support for Eq.~(\ref{entanglement}).

\section{Phenomenology}
First, we briefly summarize the phenomenology of Ref.~\cite{HO}, which should be consulted for further details.
We restrict our attention to a system of N spin 1/2's $\{\vec\sigma_i\}$ with a short-range interaction on some lattice,
with random fields that are strong enough so that all many-body eigenstates are localized.
This setup should capture much of the essence of many-body localization.

Calling the bare spins $\{\vec\sigma_i\}$ ``p-bits'' (p=physical),
we can construct ``dressed'' pseudospins $\{\vec\tau_i\}$, which are called ``l-bits'' (l=localized)~\cite{HO}.
One important criterion in building l-bits is that
all $2^N$ possible outer products of the $\tau_i^z$'s and single-spin identity operators must be constants of motion which commute with the system's Hamiltonian $H$.
By this construction, each many-body eigenstate of the Hamiltonian is one of the $2^N$ simultaneous eigenstates of all of the $\tau^z_i$'s.
This construction is always possible~\cite{lychkovskiy},
since there are in fact $(2^N)!$ possible ways of constructing a one-to-one mapping from each many-body eigenstate
to each simultaneous eigenstate of all the $\{\tau_i^z\}$.
For a given $H$ and one such mapping (which includes setting a phase for each state), one can thus define the l-bit operators $\{\vec\tau_i\}$
and expand these operators in terms of outer products of the p-bit operators $\{\vec\sigma_i\}$~\cite{HO}.  For certain random spin chains,
the existence of such a construction of localized l-bits is proven in Ref. \cite{jzi}.

For a fully many-body localized system, it is conjectured that
there exists an ``optimal'' construction of the l-bits such that the l-bits are maximally localized when they are expressed in terms of the p-bits~\cite{nonint}.
Terms in the expansion of an l-bit that involve distant p-bits have typical weights that fall off exponentially with distance.
Furthermore, we can write the Hamiltonian in terms of the l-bit operators as follows:
\begin{equation}
H = \sum_i h_i \tau_i^z + \sum_{i,j} J_{ij}\tau^z_i \tau^z_j + \sum_{i,j,k}K_{ijk}\tau^z_i \tau^z_j \tau^z_k + \ldots ~,
\end{equation}
where the typical couplings of higher-order and longer-range terms fall off exponentially.
It is then evident that the l-bits $\{\tau_i^z\}$ are localized constants of motion (a similar argument can be found in Ref.~\cite{spa2}).

We now consider the spreading of entanglement in a fully many-body localized system.
Suppose our initial state is a non-entangled pure product state of p-bits.
This is a particular linear combination of eigenstates of $H$, each of which has area-law entanglement~\cite{bauer}.
Time evolution produces area-law entanglement between nearby p-bits on a microscopic time scale.
After this early-time behavior seen in Refs.~\cite{bardarson, vosk}, the l-bit picture becomes a useful way to examine the subsequent behavior.
The l-bits become entangled because the precession rate of a given l-bit is set by its interactions with all the other l-bits.
For given fixed values of all of the other l-bits $\{\tau^z_i\}$, the typical effective interaction between two l-bits separated by a distance $x$ falls off exponentially with $x$ as $J_{eff}(x) \sim J_0 \exp(-x/\xi)$.
We use this decay to define a localization length $\xi$.
It takes roughly $t \sim 1/J_{eff}(x) \sim J_0^{-1} \exp(x/\xi)$ for this interaction to substantially affect the precession of these two l-bits and thus entangle them.
Therefore, after time $t$, the bipartite entanglement across a given ``cut'' in such a localized spin chain is due to l-bits within distance
$x \sim \xi \log (J_0 t)$ from the cut.
This explains the logarithmic growth of entanglement and the prefactor of $\xi$, the localization length.

Furthermore, we expect that in a finite system the saturation time of the entanglement entropy should thus depend on $\xi$, but be independent of the initial state provided those initial states all have the same $\xi$, which we demonstrate later. In particular,
this implies that the entanglement in two systems which possess the same localization length, but different initial conditions and therefore unequal {\it final}
entanglement entropies, should saturate at the same time. Therefore, the entanglement growth rate must also be proportional to $s_{\infty}$, where $s_{\infty}$ is the
long-time, saturated entanglement entropy per spin for a given Hamiltonian and initial state.
In the remainder of this paper, we will examine this scenario in detail.

\section{The Model and Method}
\subsection{The Model}
We consider an XXZ spin-1/2 chain with random longitudinal fields and open boundary conditions:
\begin{align}\label{hamiltonian}
H = \sum_{i=1}^{L-1} J_{\perp} (\sigma^x_i \sigma^x_{i+1} + \sigma^y_i \sigma^y_{i+1}) + J_z \sigma^z_i \sigma^z_{i+1} + \sum_{i=1}^{L} h_i \sigma^z_i ~,
\end{align}
where the
$h_i$ are static random fields uniformly drawn from $[-\eta, \eta]$.
If $J_z$ is zero, then this model is mappable to noninteracting spinless fermions with static disorder
and is thus single-particle Anderson localized.
Provided that $\eta$ is large enough and $J_z\neq 0$, this model exhibits many-body localization \cite{znidaric, pal, bardarson}.
We set the hopping strength $J_{\perp}$ = 1 to fix the energy scale
and also set $\hbar = 1$. 
We set the interaction to be $J_z = 0.2$.
Then, we study the entanglement dynamics for a range of $\eta$
and initial conditions by exactly diagonalizing the Hamiltonian
for each realization of random fields~\cite{mps}.
We average over 10 000 realizations of the random fields for $L=6$ and $L=8$, 1000 realizations for $L=10$ and $L=12$, and 100 realizations for $L=14$.
We consider only even $L$, since we study the bipartite entanglement across the midpoint of the chain.

\subsection{Method}
To compute the time dependence of the entanglement entropy, we time-evolve the spin chain from a random initial state $\ket{\Psi(0)}$.
We then partition the spin chain into two equally sized subsystems across the center bond,
between the spins at sites $\frac{L}{2}$ and $\frac{L}{2}+1$.
For each time $t$ we compute the probability operator (a.k.a. reduced density matrix) of the right half of the system, 
by tracing over all spins in the left half,
$\rho_R(t) = \mathrm{Tr}_L \{\ket{\Psi(t)}\langle\Psi(t)|\}$,
where $\ket{\Psi(t)}$ is the state of the entire spin chain.
The entanglement entropy $S(t)$ is then given by the von Neumann entropy
$S(t)=-\mathrm{Tr} (\rho_R \log_2 \rho_R) = -\mathrm{Tr}(\rho_L\log_2\rho_L)$;
note that we measure the entropy in bits.
We evaluate this quantity up to a time $t=10^{18}$,
which is long enough for the entanglement entropy to saturate for the systems considered here.

\begin{figure}
  \includegraphics[width=.5\textwidth]{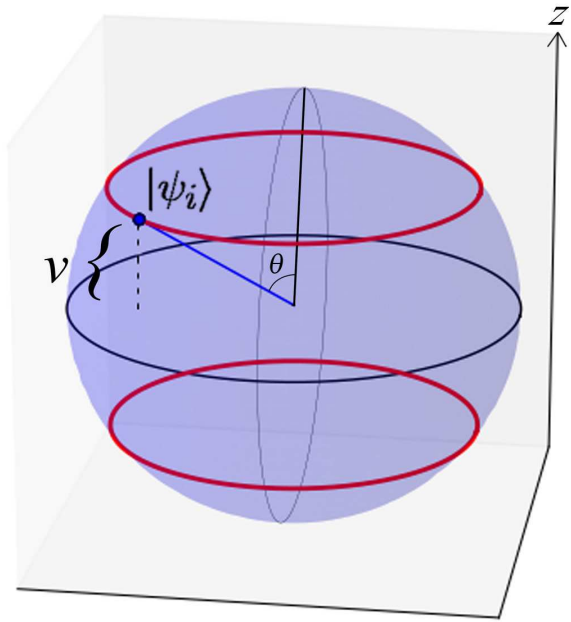}
  \caption{\label{bloch}  The initial state of each spin is chosen randomly from one of the two red circles on the Bloch sphere at a fixed height $v$ above or below the ``equator''. This parametrization allows us to control how closely the initial state resembles an eigenstate of the system and therefore the amount of dephasing and maximum entanglement entropy per spin that results.}
\end{figure}

We systematically vary the initial state of our system to investigate its effect on $S(t)$.
Each initial state is a random product state $\ket{\Psi(0)}=\bigotimes_{i=1}^L \ket{\psi_i}$,
with each spin pointing a random direction at a given height in its own Bloch sphere.
In other words, each initial state is an outer product of single-spin pure states
\begin{equation}
\ket{\psi_i}=\cos \left (\frac{\theta_i}{2} \right )\ket{\downarrow_i} + e^{i\phi_i}\sin\left(\frac{\theta_i}{2}\right)\ket{\uparrow_i} ~.
\end{equation}
As such, the initial entanglement entropy is zero.  Fig.~\ref{bloch} illustrates how each single-spin state in the product state is chosen.  $\phi_i$ is drawn
uniformly from $[0,2\pi)$ and $\cos(\theta_i)$ is chosen randomly to be either $+v$ or $-v$. Setting $v=0$, for example, yields product states of spins randomly oriented
in the $xy$ plane of the Bloch sphere,
while $v=1$ yields random product states of $\sigma^z$ eigenstates.  This parametrization is chosen so that we can generate initial states with a wide range of
saturation entanglement entropies per spin by varying $v$.  In addition, such initial states contain all possible values of the magnetization $\sum_i \sigma^z_i$,
which is a conserved quantity in the model.  For each realization of the disorder and for each $v$, 100 different initial states are evolved for all time. The entanglement entropy $S(t)$ is then computed for each $t$ and averaged over initial states.

\begin{figure}
  \includegraphics[width=.5\textwidth]{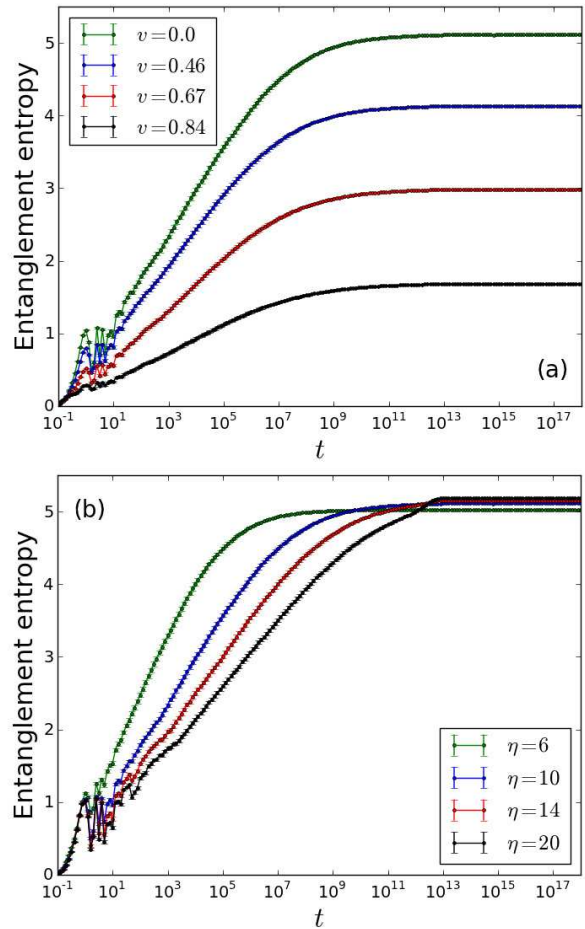}
  \caption{\label{eetime} (a) The disorder averaged entanglement entropy $S$ as a function of time, for a fixed value of disorder strength $\eta=10$ but evolved from initial states with different values of $v$. As $v$ is increased, the entanglement saturation occurs at roughly the same time, but reaches smaller values of $s_{\infty}$, leading to smaller rates of entanglement spreading. (b) $S(t)$ for different values of disorder strength $\eta$ starting from initial product states with $v=0$. As $\eta$ is increased, $S$ saturates at later times, but the saturation value is about the same, leading to smaller rates of entanglement spreading. The system size here is $L=12$.}
\end{figure}

\section{Results}
\subsection{Generic picture}
In Figs.~\ref{eetime}(a) and (b), we display the entanglement entropy $S(t)$ averaged over disorder and initial states as a function of time $t$ 
for a system of size $L=12$.
Here, as well as for all plots in this paper, error bars are shown corresponding to the 95\% confidence interval, although they may be too small to see at times.
We observe three stages: (1) A short time rapid growth and oscillation of entanglement.
This behavior mainly comes from the direct nearest-neighbor interaction between the two adjacent spins across the center bond, with a time scale set by $J_{\perp}=1$.
Since the entanglement entropy here is dominated by these two spins,
it oscillates and can be shown to be independent of system size.
(2) After the short time oscillation, many-body effects become important
and cause the entanglement entropy to grow in a logarithmic fashion.
This is the regime of interest.
The growth rate is independent of system size (not shown) and extrapolation indicates that the entanglement entropy
grows indefinitely, without bound, in the thermodynamic limit. (3) When entanglement spreads over the entire system,
finite-size effects set in and the entanglement entropy saturates.
This saturation value is proportional to the system size
and thus exhibits volume-law scaling.

\subsection{Initial configuration and final entanglement entropy per spin}

\begin{figure}
  \includegraphics[width=.5\textwidth]{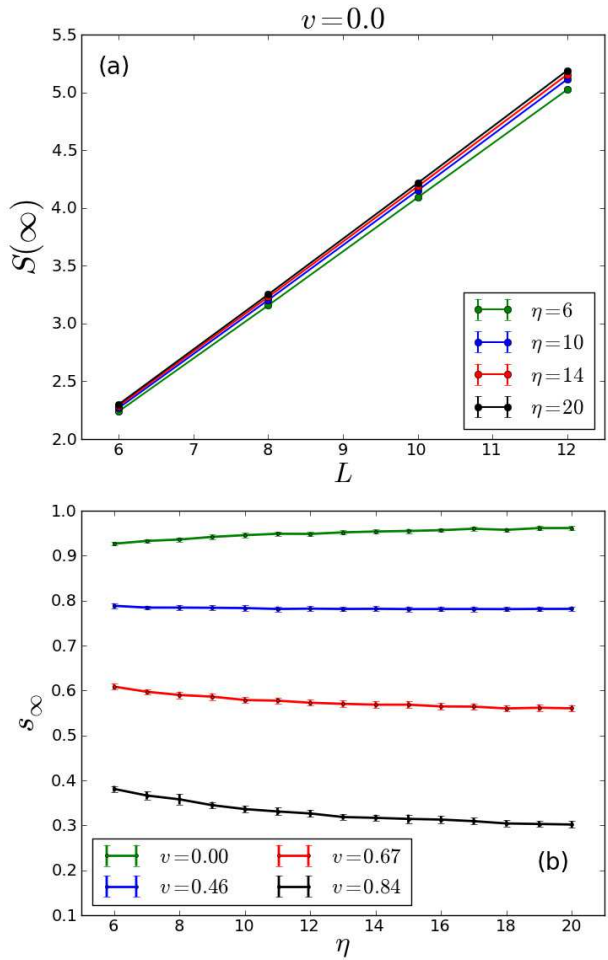}
  \caption{\label{sinf} (a) The saturated entanglement entropy, $S(\infty)$, obtained by averaging over the entanglement at large times, is shown as a function of $L$
  for $v=0$ and a few values of $\eta$.  By fitting the slopes of these lines, we estimate the saturated entanglement entropy per spin $s_{\infty}$.
  (b) The resulting $s_{\infty}$ is shown as a function of initial state $v$ and $\eta$. Small variation exists with $\eta$; however, this variation is much less than
  that due to changes in $v$.}
\end{figure}

Fig.~\ref{eetime}(a) illustrates how the spreading and the saturation value of the entanglement $S(t=\infty)$ varies with different values of $v$
for a fixed disorder $\eta=10$. Because the disorder strength is the same, the localization length $\xi$ for each of these systems should be same,
and therefore this plot corroborates our expectation that the entanglement should saturate at the same time for systems with the same localization length,
provided they are of the same size. We can also see that $S(\infty)$ is a decreasing function of $v$
(smaller for initial spins more closely aligned with the $z$-axis),
and that the rates of entanglement growth reflect this trend in saturation values,
indicating that the finite-time entanglement is proportional to the saturated value of entanglement.
This variation of the entanglement with $v$ can be understood by noting that spins in the $xy$ plane of the Bloch sphere ($v=0$)
are equal superpositions of the $\sigma^z$ eigenstates, affording them the greatest potential
for dephasing and therefore entanglement generation.
Since Refs. \cite{bardarson, pino} considered initial conditions where $v=1$,
they found a final value of entanglement smaller than what we find here for $v<1$.

To calculate the saturation values, which allow us to obtain $s_{\infty}$, we do the following:
At large times,
the two farthest apart spins have interacted, and thus the entanglement has spread over the entire system.
The bipartite entanglement entropy should then scale as~\cite{page,hyungwon}
\begin{equation}\label{entropyscaling}
S(\infty)\cong aL - b ~.
\end{equation}
The entanglement entropy per spin is $s_{\infty}=S(\infty)/(L/2)$, so $s_{\infty}= 2a - \mathcal{O}(1/L)$.
To minimize finite-size effects, we estimate $s_{\infty}$ by a linear fit to $S(\infty)$ vs. $L$ and take twice the slope. Fig.~\ref{sinf}(a) displays $S(\infty)$ vs. $L$ for four different disorder strengths at $v=0$.
The values of $s_{\infty}(v,\eta)$ obtained in this way are shown in Fig.~\ref{sinf}(b) as a function of initial state and disorder.
The strong variation of $s_{\infty}$ with $v$ contrasts with the smaller changes that occur as $\eta$ is varied.

\subsection{Localization Length}
In Fig~\ref{eetime}(b), the growth of $S$ is shown starting from initial states characterized by the same value of $v=0$,
but which are evolved according to Hamiltonians with different disorder strengths and therefore different localization lengths $\xi$.
The entanglement saturates at similar values, but the time to saturation and the spreading rate clearly vary with disorder strength, and thus $\xi$.
The localization length that enters in Eq. (1) is defined by the effective interactions between l-bits~\cite{HO}. However, we do not know how to measure this localization length directly.  Instead,
we utilize a method similar to that introduced in Ref.~\cite{pal} to estimate the localization length of the system, thus assuming that the localization length characterizing the spin correlations is proportional to the length scale of the effective l-bit interactions.

We first define the spin-spin correlation function as
\begin{equation}
\label{corr}
C_{n\alpha}^{zz}(i,j) =  \matrixel{n}{\sigma^z_{i}\sigma^z_{j}}{n}_{\alpha} - \matrixel{n}{\sigma^z_{i}}{n}_{\alpha} \matrixel{n}{\sigma^z_j}{n}_{\alpha}
\end{equation}
in eigenstate $n$ of the Hamiltonian of sample $\alpha$.

\begin{figure}
  \includegraphics[width=.5\textwidth]{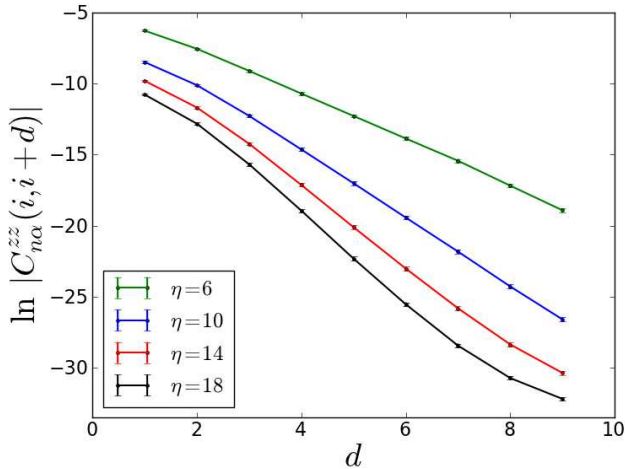}
  \caption{\label{corr} The logarithm of the correlation between centered spins separated by distance $d$, averaged over all eigenstates and over disorder realizations for $L=12$.
  By fitting the slope of these plots for $2\leq d \leq 6$, we extract a localization length $\xi$ for each value of $\eta$.
  At the very bottom of this figure these data are affected by approaching machine precision.}
 \end{figure}
\begin{figure*}[ht]
  \includegraphics[width=\textwidth]{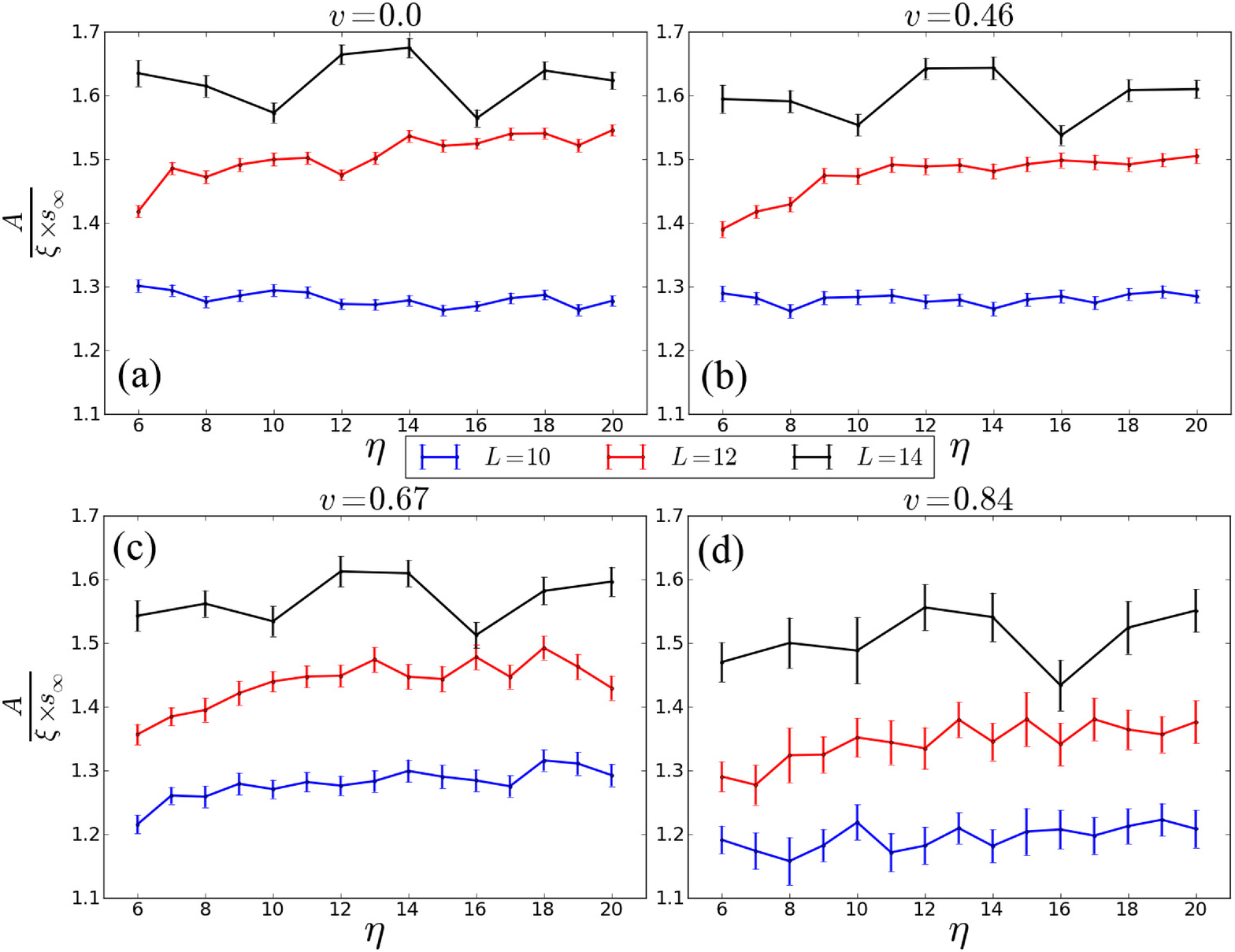}
  \caption{\label{ratios} The ratios of the spreading rate $A_{\alpha}$ to the localization length $\xi_{\alpha}$, both calculated within sample $\alpha$, divided by the saturated entanglement per spin $s_{\infty}$ and averaged over samples. The consistency of the calculated ratios between different initial conditions $v$ and also between different values of disorder strength $\eta$ is good, but finite size effects are evident, as the results differ between different values of $L$.}
\end{figure*}
In computing the correlations, we consider distances $d$ ranging from $1$ to $L-1$.
Since our model possesses open boundary conditions,
for odd values of $d$ we calculate the correlation between spins centered about the middle bond,
so that $i=(L+1-d)/2$ and $j=(L+1+d)/2$.
For even values of $d$, we measure the correlation between the two pairs of centered spins offset by one site;
i.e., between
$i=(L+2-d)/2$ and $j=(L+2+d)/2$,
and between
$i=(L-d)/2$ and $j=(L+d)/2$. The logarithm of the absolute value of the correlation, $\ln \abs{C_{n\alpha}^{zz}(d)}$, is then averaged over these two measurements. Within each sample, we then average $\ln \abs{C_{n\alpha}^{zz}(d)}$ for all $d$ over measurements in all eigenstates, and fit a line to the plot of this quantity vs. $d$ for $2\leq d\leq 6$. Fig.~\ref{corr} displays $\ln \abs{C_{n\alpha}^{zz}(d)}$ averaged over samples vs. distance $d$. The slope of this line is equal to $-1/\xi$, where $\xi$ is the localization length. It is difficult to proceed to higher values of $\eta$ because calculating $C_{n\alpha}^{zz}(d)$
for such systems calls for the subtraction of numbers beyond machine precision.
This issue also forces us to limit the values of $d$ we consider for larger $\eta$ and $L$.

\subsection{Entanglement growth rate}
We are now able to numerically test Eq.~\ref{entanglement}. We do this by first calculating the rate of logarithmic growth of $S(t)$ for each sample Hamiltonian $\alpha$ at a disorder strength $\eta$, with initial conditions corresponding to each value of $v$. This quantity will be denoted $A_{\alpha}(v)$. It is equal to the slope of the best linear fit to $S(t)$ vs. $\log{(t)}$ in the period of logarithmic growth. We then look at the ratio of this quantity to the localization length of sample $\alpha$, and perform an average of this ratio over disorder realizations. Finally, this quantity is divided by $s_{\infty}(v,\eta)$ to obtain $\langle A_{\alpha}/\xi_{\alpha}\rangle/s_{\infty}$, where the angular bracket represents disorder averaging~\cite{average}.

Figs.~\ref{ratios}(a)-(d) display plots of the resulting ratios $\langle A_\alpha/\xi_\alpha\rangle/s_{\infty}$ vs. $\eta$ for different system sizes $L$, grouped by values of $v$.
Eq.~\ref{entanglement} implies a constant value of this ratio as all parameters are varied.
The results of our numerics are mostly consistent with this expectation, with the strongest deviation being the dependence on the system size.
To begin with, there is only minimal variation of the ratio with the disorder strength $\eta$.  This probes the dependence on the localization length, and suggests that our method of estimating the localization length is reasonably reliable.

Across different values of $v$, there is good agreement between the panels of Fig.~\ref{ratios}, which checks the dependence of Eq.~\ref{entanglement} on $s_{\infty}$. The ratios calculated for $v=0.84$ are slightly lower than the rest;
we attribute this to the very limited entanglement growth that occurs as $v$ gets close to 1,
which makes the spreading rate more difficult to measure.
As already mentioned, there are sizable finite size effects in our data,
as the ratios for higher $L$ take higher values.  This difference mostly arises from finite-size effects in the estimates of $A$, not $\xi$.
We speculate that this results from the fact that boundary effects, which are felt more quickly in smaller systems,
reduce the spreading of the entanglement; the localization length, being a static quantity, is less affected.

\section{Conclusions}
We have computed the spreading of entanglement in a many-body localized system and numerically evaluated its dependence on disorder strength and initial configuration.
As we argued based on the phenomenology of localized bits, we have shown that the logarithmic growth rate of entanglement is proportional to the product of
the localization length, which is dependent primarily on disorder strength, and the final value of entanglement entropy per spin, which, to a good approximation, is dependent only on the initial configuration of the system. Within our numerical simulations, we have not seen any other independent contributions (besides finite size effects).

Our reasoning mainly relies on the existence of localized pseudospins. It is an interesting open question to find a systematic way to build these dressed pseudospins
from bare spins.  This would allow a more direct measurement of the localization length that enters into driving the spread of entanglement.

\section{Acknowledgements}
We thank Joel Moore, Mari Carmen Ba${\rm {\tilde n}}$uls, Vadim Oganesyan, and Arijeet Pal for many discussions. The calculations in this work were performed at the TIGRESS high performance computer center at Princeton University.


\begin{thebibliography}{99}

\bibitem{anderson} 
P. W. Anderson, Phys. Rev. {\bf 109}, 1492 (1958).

\bibitem{basko} 
D. Basko, I. Aleiner, and B. Altshuler, Ann. Phys. (Amsterdam) {\bf 321}, 1126 (2006).

\bibitem{vadim} 
V. Oganesyan and D. A. Huse, Phys. Rev. B {\bf 75}, 155111 (2007).

\bibitem{pal} 
A. Pal and D. A. Huse, Phys. Rev. B {\bf 82}, 174411 (2010).

\bibitem{order} 
D. A. Huse, R. Nandkishore, V. Oganesyan, A. Pal and S. L. Sondhi, Phys. Rev. B {\bf 88}, 014206 (2013).

\bibitem{pekker} 
D. Pekker, G. Refael, E. Altman, E. Demler and V. Oganesyan, arXiv:1307.3253v3.

\bibitem{bauer} 
B. Bauer and C. Nayak, J. Stat. Mech. {\bf P09005} (2013).

\bibitem{chandran} 
A. Chandran, V. Khemani, C. R. Laumann and S. L. Sondhi, Phys. Rev. B {\bf 89}, 144201 (2014).

\bibitem{znidaric} 
M. \v{Z}nidari\v{c}, T. Prosen, and P. Prelov\v{s}ek, Phys. Rev. B {\bf 77}, 064426 (2008).

\bibitem{bardarson} 
J. H. Bardarson, F. Pollmann and J. E. Moore, Phys. Rev. Lett. {\bf 109}, 017202 (2012).

\bibitem{vosk} 
R. Vosk and E. Altman, Phys. Rev. Lett. {\bf 110}, 067204 (2013) ; arXiv:1307.3256v1.

\bibitem{spa1} 
M. Serbyn, Z. Papi$\acute{\rm{c}}$ and D. A. Abanin, Phys. Rev. Lett. {\bf 110}, 260601 (2013).

\bibitem{HO} 
D. A. Huse and V. Oganesyan, arXiv:1305.4915v1.

\bibitem{pino}
M. Pino, arXiv:1403.5974v1.

\bibitem{saturation}
Strictly speaking, we can only define the final saturation value of entanglement for a finite-size subsystem.
However, the saturation entanglement entropy {\it per spin} can be defined by taking the limit of infinite time first and then the limit of
an infinitely large subsystem.


\bibitem{lychkovskiy} 
O. Lychkovskiy, Phys. Rev. A {\bf 87}, 022112 (2013).

\bibitem{jzi}
J. Z. Imbrie, arXiv:1403.7837.

\bibitem{nonint}
For a system that thermalizes, on the other hand, no such construction of localized l-bits exists.


\bibitem{spa2} 
M. Serbyn, Z. Papi$\acute{\rm{c}}$ and D. A. Abanin, Phys. Rev. Lett. {\bf 111}, 127201 (2013).

\bibitem{mps}
Although the entanglement entropy grows slowly, the method of Matrix Product States (MPS) is not favorable for our purpose,
since we need to access long times to determine the saturation value of the entanglement.
This final value of the entanglement obeys a volume-law, which requires the bond dimension to grow exponentially with the system size;
furthermore, the Trotter errors also become large at these long times.

\bibitem{page} 
D. N. Page, Phys. Rev. Lett. {\bf 71}, 1291 (1993).

\bibitem{hyungwon} 
H. Kim and D. A. Huse, Phys. Rev. Lett. {\bf 111}, 127205 (2013).

\bibitem{average}
Since the spreading rate $A$ is correlated with the localization length $\xi$,
we first take their ratio within each sample and \emph{then} perform the disorder average.
\end{thebibliography}
\end{document}